\documentclass{article}
\usepackage[utf8]{inputenc}
\usepackage[a4paper, total={6.5in, 9in}]{geometry}
\usepackage{braket}
\usepackage{amsmath,amssymb}
\usepackage{graphicx}
\newtheorem{theorem}{Theorem}

\title{Search of clustered marked states with lackadaisical quantum walks}
\author{Amit Saha$^{1*}$, Ritajit Majumdar$^2$, Debasri Saha$^1$, Amlan Chakrabarti$^1$, Susmita Sur-Kolay$^{2\dagger}$\\~\\
$^1$A. K. Choudhury School of Information Technology, University of Calcutta\\
$^2$Advanced Computing \& Microelectronics Unit, Indian Statistical Institute, Kolkata\\
Email: $^*$abamitsaha@gmail.com, $^{\dagger}$ssk@isical.ac.in}
\date{}

\begin{document}

\maketitle

\begin{abstract}
Nature of quantum walk in presence of multiple marked state has been studied by Nahimovs and Rivosh \cite{10.1007/978-3-662-49192-8_31}. They have shown that if the marked states are arranged in a $\sqrt{k} \times \sqrt{k}$ cluster in a $\sqrt{N} \times \sqrt{N}$ grid, then to find a single marked state among the multiple ones, quantum walk requires $\Omega(\sqrt{N} - \sqrt{k})$ time. In this paper, we show that using lackadaisical quantum walk with the weight of the self-loop as $\frac{4}{N(k + \lfloor{\frac{\sqrt{{k}}}{2}}\rfloor)}$, where $k$ is odd, the probability of finding a marked state increases by $\sim 0.2$. Furthermore, we show that instead of applying the quantum walk $\mathcal{O}(k)$ times to find all the marked states, classical search in the vicinity of the marked state found after the first implementation of the quantum walk can find all the marked states in $\mathcal{O}(\sqrt{k})$ time on average.
\end{abstract}

\section{Introduction}

Grover's algorithm showed that quantum computers can provide quadratic speedup for searching a marked location in an unsorted database \cite{Grover:1996:FQM:237814.237866}. While a classical algorithm requires $\mathcal{O}(N)$ time, Grover's algorithm can search the database in $\mathcal{O}(\sqrt{N})$ time. However, Benioff showed that if the $N$ data points are arranged in a $\sqrt{N} \times \sqrt{N}$ grid, then the quantum speedup is lost \cite{benioff2000space}. Since then, researches have been carried out to design faster algorithms to search an unsorted database arranged in a two or higher dimensional grid. Ambainis et al proposed an algorithm based on quantum random walk which can detect a marked state with probability $\mathcal{O}(\frac{1}{logN})$ in $\mathcal{O}(\sqrt{NlogN})$ time \cite{Ambainis:2005:CMQ:1070432.1070590}. To increase the probability, amplitude amplification is necessary, which has a time complexity of $\mathcal{O}(\sqrt{logN})$. This gives an overall running time of the algorithm to be $\mathcal{O}(\sqrt{N}logN)$. Childs and Goldstone matched this runtime with continuous time quantum walk \cite{PhysRevA.70.042312}. Ambainis et al, henceforth, proposed an algorithm which does not require amplitude amplification, and hence can perform the search in $\mathcal{O}(\sqrt{NlogN})$ time \cite{10.1007/978-3-642-35656-8_7}. Further researches have been performed to study quantum walk algorithms in other graph structures \cite{PhysRevA.67.052307, PhysRevLett.114.110503}, but in this paper we shall stick to two-dimensional grid.\\

Most of the quantum walk based search algorithms consider one or two marked locations. In \cite{10.1007/978-3-662-49192-8_31}, Nahimovs and Rivosh considered searching multiple marked states in a $\sqrt{N} \times \sqrt{N}$ grid. They showed that if $k$ marked states are grouped in a $\sqrt{k} \times \sqrt{k}$ block, then the algorithm of \cite{Ambainis:2005:CMQ:1070432.1070590} can perform the search in $\Omega(\sqrt{N} - \sqrt{k})$ time. Whereas if the $k$ marked locations are distributed uniformly over the grid, then the algorithm requires $\mathcal{O}(\sqrt{\frac{N}{k}log\frac{N}{k}})$ time. In \cite{978-3-319-29817-7}, They also showed that when $k$ is even, quantum walk exceptionally fails to find any of the marked locations. Wong proposed lackadaisical quantum walk \cite{Wong2018}, where each location on the grid contains $l$ number of self loops. These loops give some probability to the walker to remain at its location, hence making the walk lazy. In \cite{wang2017adjustable}, the authors have used adjustable self-loops and have shown that when the weight of the self-loop is $\frac{4}{N}$, the probability of finding a single marked state in a $\sqrt{N} \times \sqrt{N}$ grid increases, as compared to normal quantum walk (not lackadaisical).\\

In this paper we extend the model of \cite{wang2017adjustable} to multiple marked states arranged in a $\sqrt{k} \times \sqrt{k}$ cluster, where $k$ is odd. We show by simulation that adjusting the weight of the self-loop as $\frac{4}{N(k + \lfloor{\frac{\sqrt{{k}}}{2}}\rfloor)}$, the probability of finding the states increases by $\sim 0.2$ for all values of $N$. In Table~\ref{tab:summary} we provide an overview of the time complexity and success probability of both classical and quantum random walk algorithms to detect single and multiple marked states. We show by simulation that using weight of self loop as $\frac{4}{N(k + \lfloor{\frac{\sqrt{{k}}}{2}}\rfloor)}$, the number of steps required is less than that of quantum walk with no self loop.

\begin{table}[h!]
\caption{Time complexity \& success probability of finding single and multiple marked states with existing algorithms}
\centering
\begin{tabular}{ |c|cc|cc| }
  \hline
& \multicolumn{2}{|c}{Single Mark State}&\multicolumn{2}{|c|}{Clustered Mark States}\\
\hline
 & Success Probability & Steps & Success Probability & Steps \\ \hline
Random Walk & $\mathcal{O}(\frac{1}{N})$ & $\mathcal{O}(NlogN)$ & $\mathcal{O}(\frac{k}{N})$ & $\mathcal{O}(\frac{N}{k}log\frac{N}{k})$ \\
Quantum Walk & $\mathcal{O}(\frac{1}{logN})$ & $\mathcal{O}(\sqrt{NlogN})$ & $\mathcal{O}(\frac{1}{log(N/k)})$ & $\Omega(\sqrt{N} - \sqrt{k})$ \\ 
 \hline
\end{tabular}
\label{tab:summary}
\end{table}

\section{Discrete time quantum walk}
A quantum random walk consists of position Hilbert Space $H_p$ and coin Hilbert Space $H_c$. A quantum state consists of these two degrees of freedom, $\ket{c} \otimes \ket{v}$. where $\ket{c} \in H_c$ and $\ket{v} \in H_p$. A step in quantum walk is a unitary evolution $U = S.(C \otimes I)$ where $S$ is the shift operator and $C$ is the coin operator, which acts only on the coin Hilbert Space $H_c$. If we consider a $\sqrt{N} \times \sqrt{N}$ grid, then the quantum walk starts in a superposition
\begin{equation}
    \ket{\psi(0)} = \frac{1}{\sqrt{4N}}(\sum_{i=1}^{4}\ket{i} \otimes \sum_{x,y=1}^{\sqrt{N}}\ket{x,y})
\end{equation}

where each location $(x,y)$ corresponds to a quantum register $\ket{x,y}$ with $x,y \in \{1, 2, \hdots, \sqrt{N}\}$ and the coin register $\ket{i}$ with $i \in \{\leftarrow, \rightarrow, \uparrow, \downarrow\}$. The most often used transformation on the coin register is the Grover's Diffusion Transformation $D$
\begin{center}
    $D = \frac{1}{2}\begin{pmatrix}
    -1 & 1 & 1 & 1\\
    1 & -1 & 1 & 1\\
    1 & 1 & -1 & 1\\
    1 & 1 & 1 & -1
    \end{pmatrix}$
\end{center}

The Diffusion Operator can also be written as $D = 2\ket{s_D}\bra{s_D} - I_4$, where $\ket{s_D} = \frac{1}{\sqrt{4}}\sum_{i=1}^{4}\ket{i}$.\\

The transformation creates a superposition of the coin states $\ket{i}$, which in turn governs the shift operation. In this paper, we apply the Flip-Flop Shift transformation S proposed in \cite{Ambainis:2005:CMQ:1070432.1070590}

\begin{center}
    $\ket{i,j,\uparrow} = \ket{i,j-1,\downarrow}$\\
    $\ket{i,j,\downarrow} = \ket{i,j+1,\uparrow}$\\
    $\ket{i,j,\leftarrow} = \ket{i-1,j,\rightarrow}$\\
    $\ket{i,j,\rightarrow} = \ket{i+1,j,\leftarrow}$
\end{center}

It is easy to see that $\ket{\psi(t)}$ is a $+1$ eigenstate of the operator $U = S.(D \otimes I)$. A perturbation is created in the quantum state by applying the coin operator $-I$ instead of $D$ for marked locations. A general quantum walk algorithm applies this unitary operation (appropriately for the marked and the unmarked states) $t$ times to create the state $\ket{\psi(t)}$ such that $\braket{\psi(t)|\psi(0)}$ becomes close to 0. Measurement of the state $\ket{\psi(t)}$ is expected to give the marked location with high probability.

\subsection{Lackadaisical quantum walk}

In lackadaisical quantum walk, the coin degree of freedom is five-dimensional, i.e. $i \in \{\leftarrow, \rightarrow, \uparrow, \downarrow, . \}$. The flip-flop transformation conditioned on the $\ket{.}$ coin state is

\begin{center}
    $S(\ket{i,j} \otimes \ket{.}) = \ket{i,j} \otimes \ket{.}$
\end{center}

If $l$ self-loops are allowed, then the Coin operator will be $D = 2\ket{s_D}\bra{s_D} - I_5$, where 

\begin{center}
    $\ket{s_D} = \frac{1}{\sqrt{4+l}}(\ket{\uparrow} + \ket{\downarrow} + \ket{\leftarrow} + \ket{\rightarrow} + \sqrt{l}\ket{.})$
\end{center}

In \cite{Wong2018}, Wong showed that using lackadaisical quantum walk with $l = \frac{4}{N}$, the success probability of finding a marked state in an $\sqrt{N} \times \sqrt{N}$ grid becomes close to 1 in $\mathcal{O}(\sqrt{NlogN})$ time.

\section{Clustered marked state over 2-D grid}
\begin{figure}[h!]
    \centering
    \includegraphics[scale=0.4]{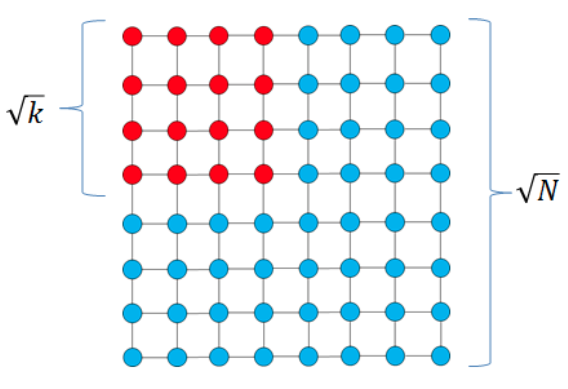}
    \caption{Grouped placement of $k$ marked locations \cite{10.1007/978-3-662-49192-8_31}}
    \label{fig:cluster}
\end{figure}

In \cite{10.1007/978-3-662-49192-8_31}, the authors have considered multiple marked locations. In accordance with the paper, we consider the walk is taking place on a $\sqrt{N} \times \sqrt{N}$ grid and there are $k$ marked locations arranged in a $\sqrt{k} \times \sqrt{k}$ cluster (shown in Fig.~\ref{fig:cluster}).

In \cite{10.1007/978-3-662-49192-8_31}, the authors have used quantum walk where the weight on the self loop is $0$, or in other words, is not lackadaisical. In \cite{wang2017adjustable}, the authors have shown that using a weight of $\frac{4}{N}$ for the self loop, the probability to find a single marked state increases with respect to non-lackadaisical walk. However, for multiple marked states, this weight provides a probability poorer than non-lackadaisical walk.\\

Changing the weight of the self loop to be $\frac{1}{4N}$, the total probability becomes close to $0.85$ in $\mathcal{O}(\sqrt{N})$ steps for $\sqrt{N} > 12$. The probability of finding each marked state is \emph{total probability/$k$}. The total probability exceeds the probability in non-lackadaisical walk by $\sim 0.2$. However, if the weight on the self loop is a particular function of both the size of the grid ($N$) and the number of marked states ($k$), then the probability is shown to be even higher.\\

\begin{figure}
    \centering
    \includegraphics[scale=0.7]{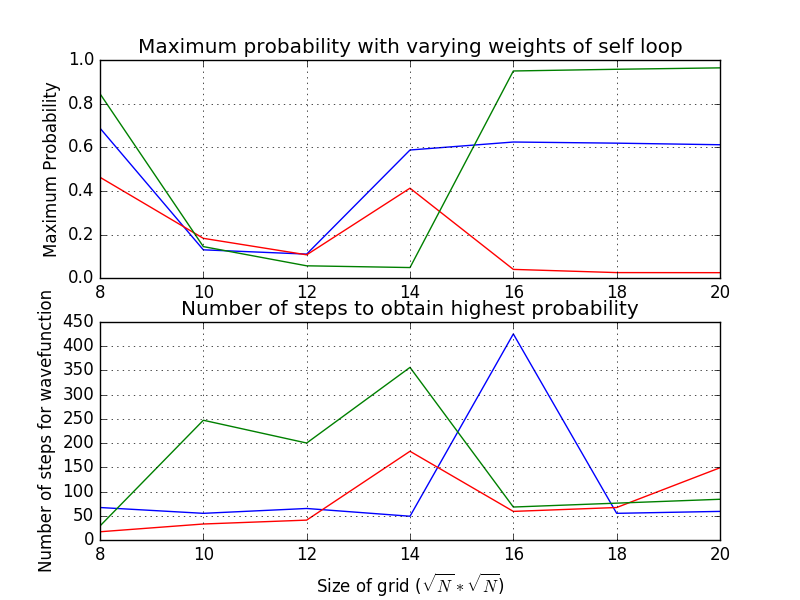}
    \includegraphics[scale=0.7]{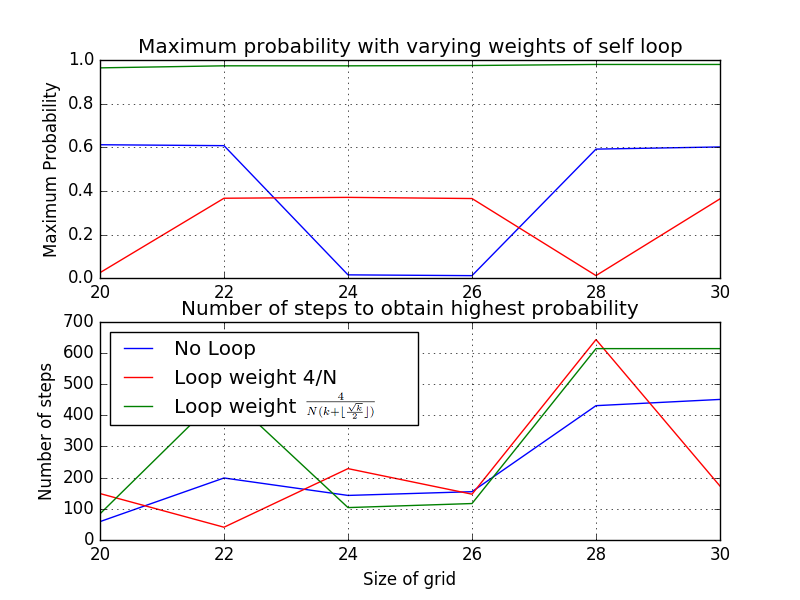}
    \caption{Probability and number of steps for $8 \le \sqrt{N} \le 30$}
    \label{fig:test1}
\end{figure}

We set the probability of the self loop as $\frac{4}{N(k + \lfloor{\frac{\sqrt{{k}}}{2}}\rfloor)}$ for odd values of $k$. Our simulations show that using this weight, the probability is close to $0.8$ for $\sqrt{N} \le 14$ and exceeds $0.95$ for larger values of $N$. In Figure~\ref{fig:test1} and \ref{fig:test2}, we represent the results of our simulations graphically. In each Figure, we have shown three cases - (i) weight of self loop is 0 (blue color) \cite{10.1007/978-3-662-49192-8_31}, (ii) weight of self loop is $\frac{4}{N}$ (red color) \cite{wang2017adjustable} and (iii) weight of self loop is $\frac{4}{N(k + \lfloor{\frac{\sqrt{{k}}}{2}}\rfloor)}$ (proposed weight, green color). We show both the highest total probability and the number of steps required to attain this probability. We show the results of simulations for $k = 9$, i.e. 9 marked states arranged in a $3 \times 3$ cluster. The grid size, in the graphical representations, is varied from $\sqrt{N} = 8$ to $30$. For brevity, the plots show the probability to find the marked state for even values of $\sqrt{N}$ only. The odd values are exactly similar.

\section{Local search in vicinity}

The original paper by Nahimovs and Rivosh \cite{10.1007/978-3-662-49192-8_31} considered finding only one marked state among the multiple ones. In this section, we consider the scenario where it is required to find all the marked states. This may be necessary for scenarios like tracking an object in a picture. An obvious choice is to run the quantum walk algorithm multiple times to find all the marked states. Since the probability of finding each of the marked states is equal, and there are $k$ marked states, the algorithm needs to be run $\mathcal{O}(k)$ times. However, in current scenario, it is more costly to apply quantum walk $\mathcal{O}(k)$ times than to apply classical search for the similar times. We show next that applying classical search in the vicinity of the marked state require $\mathcal{O}(\sqrt{k})$ times to find all the marked states.\\

It is important to consider that the starting point is a marked state of some coordinate $(l,m)$ where $l$ and $m$ are unknown. Since, the structure of the marked states is always a grid, and the number of marked states are also known a-priori, one can search in the vicinity of the $(l,m)$ point to find its relative location with respect to the boundary of the grid of marked states. Thus by knowning the values of $l$ and $m$, the position of all the marked states can be obtained in $\mathcal{O}(1)$ time.

\begin{theorem}
Starting from a marked state, chosen uniformly from a $\sqrt{k} \times \sqrt{k}$ cluster, the average time to find all the marked states in that cluster, using classical search, is $\mathcal{O}(\sqrt{k})$.
\end{theorem}

\textbf{Proof} Let us consider that the lower left marked state in the $\sqrt{k} \times \sqrt{k}$ cluster has a coordinate $(0,0)$. The search algorithm starts from some point $(l,m)$ on the grid, where $l$ and $m$ are not known a priori. Let us assume that the classical search algorithm always moves once towards the left and and then down from the starting point $(l,m)$ until it hits the boundary. To determine the lower boundary, it has to move $l+1$ steps and to determine the left boundary, it has to move $m+1$ steps. The extra step in both cases is required to determine that the boundary has indeed been reached.\\

Consider the leftward walk of the search algorithm. In the best case, when the starting point is on the left-most boundary, a single step is sufficient. Whereas, if the starting point is on the right-most boundary, then $\sqrt{k}+1$ steps are necessary. Since, each location on the grid is equally likely to start with, and for the horizontal walk, there are $\sqrt{k}$ possible starting points, the expected time requirement is

\begin{equation*}
    \frac{1 + 2 + ... + (\sqrt{k}+1)}{\sqrt{k}} = \frac{(\sqrt{k}+1)(\sqrt{k}+2)}{\sqrt{k}} = \mathcal{O}(\sqrt{k})
\end{equation*}

A similar approach is required for the downward movement from the lasting position, which, by symmetry will also require $\mathcal{O}(\sqrt{k})$ time. Hence, the total time requirement to search for the relative values of $l$ and $m$ with respect to the grid of marked states is $\mathcal{O}(\sqrt{k})$.\\

From the above theorem, it is evident that applying quantum walk once, and then search using classical algorithm can find all the marked states in $\mathcal{O}(\sqrt{k})$ times, whereas using only quantum walk would have required $\mathcal{O}(k)$ time.

\section{Conclusion}

In this paper we have studied the application of quantum random walk on an $\sqrt{N} \times \sqrt{N}$ grid for all odd values of $k \ge 1$ marked states, where the marked states are arranged in a $\sqrt{k} \times \sqrt{k}$ cluster. Our simulations show that using lackadaisical quantum walk, where the weight of the self loop is $\frac{4}{N(k + \lfloor{\frac{\sqrt{{k}}}{2}}\rfloor)}$, the probability of finding a marked state becomes close to 1 in time less than that of quantum walk with no self loop. Classical search in the vicinity can find all the marked states in $\mathcal{O}(\sqrt{k})$ time on average. Hence, instead of using quantum walk for $\mathcal{O}(k)$ times, one can use classical search after a single step of quantum walk and still obtain the speedup. Future scopes of this work is to prove this bound mathematically, and to find the optimum weight for which the highest probability is attained.

\bibliographystyle{plain}
\bibliography{sample}

\end{document}